\documentclass[12pt]{article}
\usepackage[T1]{fontenc}
\usepackage[numbers,sort&compress]{natbib}
\usepackage{lmodern,microtype,amsmath,amssymb,mathtools,bm}
\usepackage[margin=0.88in]{geometry}
\usepackage{graphicx,booktabs,array,placeins,needspace}
\usepackage{doi}
\usepackage{float}
\usepackage{orcidlink}

\usepackage[dvipsnames]{xcolor}
\definecolor{linkblue}{RGB}{0,72,150}
\definecolor{citegreen}{RGB}{0,110,70}
\definecolor{urlpurple}{RGB}{105,55,135}
\usepackage[font=small,labelfont=bf]{caption}
\usepackage{hyperref}
\hypersetup{
  colorlinks=true,
  linkcolor=linkblue,
  citecolor=citegreen,
  urlcolor=urlpurple,
  pdftitle={Entanglement as a resource for self-organization and stabilization},
  pdfauthor={Kazuki Ikeda},
  pdfsubject={},
  pdfkeywords={},
  pdfcreator={},
  pdfproducer={}
}
\AtBeginDocument{
  \hypersetup{
    colorlinks=true,
    linkcolor=linkblue,
    citecolor=citegreen,
    urlcolor=urlpurple
  }
}

\usepackage[nameinlink,noabbrev,capitalise]{cleveref}

\graphicspath{{../figures/}}
\newcommand{\D}{\mathcal D}
\newcommand{\Lind}{\mathcal L}
\newcommand{\PB}{\mathcal P_{\mathcal B}}

\newcommand{\Tr}{\operatorname{Tr}}
\newcommand{\ee}{\varepsilon}
\newcommand{\ellinf}{\ell_\infty}
\newcommand{\einf}{e_\infty}
\newcommand{\leak}{\kappa_{\mathrm l}}
\newcommand{\repair}{\kappa_{\mathrm r}}
\newcommand{\Db}{D_{\mathrm{bg}}}
\newcommand{\Dcl}{D_{\mathrm{cl}}}
\newcommand{\Bsep}{B_{\mathrm{sep}}}
\title{Quantum entanglement as a resource for self-organization and stabilization}
\author{Kazuki Ikeda\,\orcidlink{0000-0003-3821-2669}\\\small University of Massachusetts Boston}
\date{}
\begin{document}
\maketitle
\begin{abstract}
Can quantum entanglement generate macroscopic order that persists after the entanglement itself disappears? We establish such a mechanism in a driven-dissipative bosonic lattice, where local entanglement enables transport that amplifies spatial fluctuations into extended Turing patterns. For the full nonlinear dynamics with noise and disorder, we derive rigorous finite-time bounds on pattern intensity for all admissible separable evolutions, allowing arbitrary classical correlations. Quantum dynamics produces pattern intensities beyond these bounds during both formation and stabilization under noise. The excess intensity also provides a lower bound on the accumulated transport contribution beyond the separable limit. We further introduce an environment that responds to emerging order by suppressing local coherence and activating classical transport. The patterns then persist as local entanglement is lost, with both stages arising from a single time-independent Lindblad equation. These results show how transient quantum resources can establish persistent macroscopic order, opening new directions for harnessing self-organization in nonequilibrium systems.
\end{abstract}

\section{Introduction}

A homogeneous state can develop spatial structure when it becomes unstable to perturbations at a finite wave number. In the Turing mechanism, transport destabilizes a state that is stable under local reactions alone \cite{Turing1952}. Local activation combined with inhibition over a longer range provides one physical realization \cite{GiererMeinhardt1972}. The instability selects a spatial scale, while nonlinear saturation, mode competition and anisotropy determine the emerging morphology \cite{CrossHohenberg1993}. Understanding a pattern therefore requires explaining both its formation and the dynamics that sustain it.

These questions connect reaction--diffusion theory with self-organization in driven-dissipative systems \cite{MainiPainterChau1997,KondoMiura2010,PrigogineLefever1968,Haken1975}. Chemical experiments exhibit transitions from uniform states to hexagonal and striped patterns, while biological pigmentation patterns can reorganize after local disruption \cite{OuyangSwinney1991,YamaguchiYoshimotoKondo2007}. Extensions to heterogeneous media and networks broaden the range of transport processes that can produce spatial order \cite{KrauseNearEquilibrium2021}. Related mechanisms operate in optical cavities and atomic gases, where coherent driving and feedback generate spatial structure \cite{LugiatoLefever1987,DomokosRitsch2002,BaumannEtAl2010}.

Cosmology offers a broader perspective on the relation between the origin of structure and its subsequent evolution. Gravitational instability amplifies primordial inhomogeneities \cite{osti_4158113,Bond1996,Aschwanden2018}, while inflationary models connect these inhomogeneities to quantum fluctuations and their later classical description \cite{MUKHANOV1992203,Polarski:1995jg}.

Autonomous stabilization provides a complementary perspective on how organized states are reached and maintained. In distributed computing, self-stabilization concerns convergence from arbitrary configurations to valid states and their subsequent persistence \cite{10.1145/361179.361202,10.7551/mitpress/6156.001.0001}. In quantum systems, engineered dissipation enables autonomous preparation and stabilization of quantum states \cite{PhysRevA.78.042307,Verstraete2009,Shankar2013,Reiter2017}. These perspectives motivate a common question about the resources needed to establish collective organization and those needed to preserve it as the system and its environment evolve.

Quantum effects on spatial organization have been investigated in nonlinear optical systems and cavity-mediated atomic reordering \cite{gatti,maschler}. Symmetry-breaking transitions to inhomogeneous states have been studied in coupled quantum oscillators \cite{BandyopadhyayKhatunBanerjee2021,paul}, while Turing instability in quantum activator–inhibitor systems has been shown to generate nonuniformity and entanglement in two coupled units \cite{KatoNakao2022}. Subsequent work explored mode competition and pattern selection in a one-dimensional bosonic chain \cite{ComparatoGarganoLoFranco2026}. Two-dimensional quantum Turing patterns were explicitly constructed in \cite{2026arXiv260726331I}, followed by a systematic study of their morphologies and the relationship between spatial order and entanglement witnesses \cite{2026arXiv260820151I}. Related experiments have demonstrated pattern formation and stabilization in driven quantum gases \cite{zhang,liebster}.

These developments raise two related questions: 
\begin{enumerate}
    \item \textbf{Necessity of entanglement.}
    Under what physical conditions is entanglement necessary for the formation and stabilization of macroscopic spatial order?

    \item \textbf{Autonomous classicalization.}
    Can order formed through quantum dynamics persist when interactions with the environment remove the entanglement involved in its formation?
\end{enumerate}

We address both questions in a driven-dissipative bosonic lattice with coherence-dependent transport, extended to include an environment that responds to the emerging pattern. Our main results are as follows.

\begin{itemize}
\item \textbf{Local entanglement and global spatial order.} Entanglement between two bosonic modes within each lattice cell enables spatial patterns extending across the entire system, even though correlations between different cells remain classical [Sec. \ref{sec:coherence}, Eqs. \eqref{eq:eln}--\eqref{eq:causal-threshold}, Figs. \ref{fig:geometry-overview}, \ref{fig:resource-geometry}(a)].
\item \textbf{Rigorous bounds without entanglement.} We derive upper bounds on the intensity of spatial patterns attainable by any admissible evolution that remains separable, even in the presence of arbitrary classical correlations. These bounds provide quantitative criteria for identifying spatial order that requires entanglement [Sec. \ref{sec:separable-bounds}, Eqs. \eqref{eq:Gbound}--\eqref{eq:conditional-maint}].
\item \textbf{A quantitative link between entanglement and transport.} The excess pattern intensity provides a lower bound on the accumulated transport contribution beyond what separable states can supply [Sec. \ref{sec:resource-definition}, Eq. \eqref{eq:geometry-transport-budget}, Fig. \ref{fig:resource-geometry}(b)].
\item \textbf{Formation and stabilization beyond these bounds.} Quantum dynamics generates Turing patterns whose mean intensities exceed the separable bounds during both pattern formation and subsequent evolution under noise. This establishes that entanglement is necessary to attain and sustain the observed pattern intensities [Figs. \ref{fig:geometry-overview}, \ref{fig:resource-geometry}(c) and \ref{fig:geometry-maintenance}(c,d)].
\item \textbf{Persistence after autonomous classicalization.} Environmental transitions suppress local coherence generation and activate classical transport, allowing the established order to persist as local entanglement vanishes. Both stages arise from the same time-independent dynamics
[Sec.~\ref{sec:Classicalization}, Eqs.~\eqref{eq:classicalization_transport} and~\eqref{eq:classicalization_jump}, Fig.~\ref{fig:classicalization}].
\end{itemize}

These results make the need for entanglement in macroscopic self-organization testable through collective observables under specified dynamical constraints. The environmental extension further shows that the resources required to establish order can differ from those sufficient to sustain it within a single autonomous system. This opens a route to determining how the amount, spatial distribution and lifetime of entanglement constrain collective organization. Extending this perspective to other collective phenomena could guide the design of autonomous nonequilibrium systems that use transient quantum resources to generate structures sustained by classical dynamics.

\section{Model and instability}\label{sec:model}
We study a lattice of bosonic dimers in which intermode coherence controls the transport that amplifies spatial variations. The total occupation of each cell defines the local field, and the dynamics is governed by a time-independent Lindblad equation.
\subsection{Bosonic lattice dynamics}\label{sec:local-space}
We consider a rectangular array $\Lambda$ of $M$ dimer cells on a square lattice, with periodic boundaries and unit spacing. Each cell contains two bosonic modes $a_x,b_x$, with number operators $n_{a,x}=a_x^\dagger a_x$ and $n_{b,x}=b_x^\dagger b_x$. The two imbalanced number configurations differ only in which mode carries one additional boson. Their coherent superpositions control local transport, while a common shift of both occupations changes the pattern amplitude. In the joint number basis, set
\begin{align}
 |n,L\rangle&=|n,n+1\rangle, &|n,R\rangle&=|n+1,n\rangle,\\
 |n,\pm\rangle&=(|n,L\rangle\pm|n,R\rangle)/\sqrt2,
 &|n,0\rangle&=|n,n\rangle, \qquad 0\le n\le n_{\max}. \label{eq:space}
\end{align}
The labels $s=+,-,0$ distinguish the two relative-phase states and the equal-occupation leakage state. The occupation contrast $\hat u_x$ describes the spatial pattern, $C_x$ measures intermode coherence, and $P_{0,x}$ projects onto the leakage sector. Define
\begin{align}
 \hat u_x&=\ee\left[(n_{a,x}+n_{b,x}-1)/2-n_0\right], & n_0&=U/\ee,\quad n_{\max}=2U/\ee,\\
 C_x&=\sum_n\bigl(|n,L\rangle\langle n,R|+\mathrm{h.c.}\bigr),
 &P_{0,x}&=\sum_{n=0}^{n_{\max}+1}|n,n\rangle\langle n,n|.
\end{align}
Here $\ee>0$ is the contrast step under a common number shift, and $U>0$ sets the occupation cutoff. We choose $n_{\max}=2U/\ee$ to be an integer. The operator $C_x$ has eigenvalues $+1,-1,0$, and $|u_x|\le U_e:=U+\ee/2$. The contrast is $u=\ee(n-n_0)$ for $s=\pm$ and $u=\ee(n-n_0-1/2)$ for $s=0$. The auxiliary projector onto $|n_{\max}+1,n_{\max}+1\rangle$ ensures that the partial transpose of $2P_{0,x}-C_x$ is positive, as shown in Appendix~\ref{app:weighted-derivation}. This state remains unoccupied throughout the evolution considered here.

We use the Lindblad dissipator $\D[A]\rho=A\rho A^\dagger-\tfrac12\{A^\dagger A,\rho\}$. The operators $T_x^\pm$ increase or decrease both occupation numbers by one, preserve $s$, and vanish at the corresponding occupation cutoff. The rate $r$ controls the generation of intermode coherence, while $\eta$ sets the strength of local number dephasing. Coherence generation and number dephasing are described by
\begin{equation}
 R_x=\sum_n|n,+\rangle\langle n,-|,\qquad
 \Lind_{\mathrm{ph}}=\sum_x\left[r\D[R_x]+2\eta\D[n_{a,x}]+2\eta\D[n_{b,x}]\right].
\end{equation}
Leakage transfers a cell from either imbalanced occupation sector to the equal-occupation sector at rate $\leak$. Repair returns it to the two imbalanced sectors with equal probabilities and total rate $\repair$. These channels selectively remove and restore one physical particle. Their jump operators $V_{x,\sigma}$ and $W_{x,\sigma}$ are specified in Appendix~\ref{app:model-details}.

The field dynamics couples neighboring cells through coherence-dependent transport. In the following operator expressions, $u_x$ denotes the occupation-contrast operator $\hat u_x$. Let $L$ be the positive nearest-neighbor graph Laplacian, $(Lu)_x=\sum_{y\sim x}(u_x-u_y)$, and let $L_y$ contain the contributions from its two vertical neighbors. The coefficient $D_2$ sets the transport strength and $w$ the width of the static disorder. Throughout, $\delta,D_2>0$, $\Dcl\ge0$ and $0\le w<1$. The onsite damping and symmetric positive edge weights obey
\begin{equation}
 \delta_x\in[\delta_-,\delta_+],\quad\delta_\pm=\delta(1\pm w),\qquad
 j_{xy}=j_{yx}\in[D_2(1-w),D_2(1+w)].
\end{equation}
Write $J=D_2(1+w)$. The coherence-dependent transport operator is
\begin{equation}
 (L_{j,C}u)_x=\sum_{y\sim x}j_{xy}\frac{C_x+C_y}{2}(u_x-u_y).
\end{equation}
The drift combines local damping, suppression of short-wavelength variations, anisotropy, coherence-dependent transport and nonlinear saturation,
\begin{equation}
 F_x=-\delta_xu_x-D_4(L^2u)_x-\zeta(L_yu)_x+(L_{j,C}u)_x
      +\Dcl(Lu)_x+\nu u_x^2-bu_x^3-cu_x^5. \label{eq:F}
\end{equation}
Here $D_4,b>0$ and $c,\zeta\ge0$. The quadratic term $\nu u_x^2$ breaks the symmetry between positive and negative contrast. The coefficient $\Dcl$ allows weak gradient amplification independent of coherence. This contribution and the static disorder are included in the comparison bound.

The field transition rates use a positive smoothing parameter $s_0$ and background occupation-noise strength $\Db$,
\begin{equation}
 q_x^\pm=\frac{\sqrt{F_x^2+s_0^2}\pm F_x}{2\ee},\qquad
 J_x^\pm=T_x^\pm\sqrt{q_x^\pm},\qquad
 B_x^\pm=\frac{\sqrt{\Db}}{\ee}T_x^\pm. \label{eq:fieldjumps}
\end{equation}
Since the occupation-contrast operators $\hat u_x$ commute with the coherence operators $C_y$, the rate operators are defined in their common eigenbasis. The operator ordering ensures that each rate is evaluated before the number shift. The parameter $s_0$ smooths the rates at $F_x=0$. The background channel supplies additive occupation noise. The density operator $\rho$ obeys $\dot\rho=\Lind\rho$, with generator
\begin{equation}
 \Lind=\Lind_{\mathrm{ph}}+
 \sum_{x,\sigma=L,R}\bigl(\D[V_{x,\sigma}]+\D[W_{x,\sigma}]\bigr)
 +\sum_{x,\sigma=\pm}\bigl(\D[J_x^\sigma]+\D[B_x^\sigma]\bigr). \label{eq:gen}
\end{equation}
Each jump operator acts only on cells within two lattice spacings of its central cell. 

Away from the occupation cutoffs, the field jumps in Eq.~\eqref{eq:fieldjumps} contribute the conditional moments
\begin{equation}
 \mathbb E[du_x\mid u,C]/dt=F_x,\qquad
 \mathbb E[(du_x)^2\mid u,C]/dt=\ee\sqrt{F_x^2+s_0^2}+2\Db. \label{eq:moments}
\end{equation}

\subsection{Coherence, entanglement and the Turing instability}\label{sec:coherence}
All field shifts preserve $C_x$ and $P_{0,x}$. At one cell, let $e=\langle C_x\rangle$ be the coherence and $\ell=\langle P_{0,x}\rangle$ the leakage probability. For the dynamics in Eq.~\eqref{eq:gen}, the unconditional expectation values $e$ and $\ell$ obey the closed equations
\begin{equation}
 \dot\ell=\leak(1-\ell)-\repair\ell,\qquad
 \dot e=r(1-\ell)-(r+2\eta+\leak)e. \label{eq:source}
\end{equation}
We take $r,\eta,\leak,\repair\ge0$ and assume positive relaxation rates $k=\leak+\repair>0$ and $a=r+2\eta+\leak>0$. Their stationary values are
\begin{equation}
 \ellinf=\frac{\leak}{\leak+\repair},\qquad
 \einf=\frac{r(1-\ellinf)}{r+2\eta+\leak}.
 \label{eq:sourceinf}
\end{equation}
Local coherence approaches its stationary value before substantial spatial growth occurs [Fig.~\ref{fig:resource-geometry}(a)]. The closed equations for $e$ and $\ell$ remain valid with field noise and static disorder. Appendix~\ref{app:source-transient} gives the exact transient solution.

Separability across the two physical modes requires positivity under partial transpose (PPT) \cite{peres}. For the local occupation space in Sec.~\ref{sec:local-space}, every PPT state obeys $e\le2\ell$, as derived in Appendix~\ref{app:weighted-derivation}. Thus $e>2\ell$ certifies local intermode entanglement.

\Needspace{7\baselineskip}
The negativity $\mathcal N$ is the sum of the absolute values of the negative eigenvalues of the partially transposed two-mode state. With logarithmic negativity $E_{\mathrm{LN}}=\log_2(1+2\mathcal N)$ and $[z]_+=\max(z,0)$, the corresponding bounds on the unconditional two-mode state are
\begin{equation}
 \mathcal N\ge\frac{[e-2\ell]_+}{4},\qquad
 E_{\mathrm{LN}}\ge\log_2\left(1+\frac{[e-2\ell]_+}{2}\right). \label{eq:eln}
\end{equation}
For the initial states considered here, the Lindblad dynamics preserves density matrices diagonal in the joint $|n,s\rangle$ basis. The populations in this basis therefore evolve according to a continuous-time Markov process, which can be sampled exactly. Although the basis states $|n,\pm\rangle$ are entangled across the two physical modes, their mixtures need not be entangled. We certify entanglement of the unconditional local state through $e>2\ell$ and the bounds in Eq.~\eqref{eq:eln}.

The separability constraint and the instability threshold can now be compared directly. In the homogeneous deterministic limit with rapid phase relaxation, we replace the local coherence in the linear drift by its uniform mean $e$. A Fourier perturbation about $u=0$ then has growth rate
\begin{equation}
 \sigma(\bm k;e)=-\delta+(D_2e+\Dcl)\omega(\bm k)-D_4\omega(\bm k)^2-\zeta\omega_y(\bm k),
 \label{eq:linear-guide}
\end{equation}
where $\omega(\bm k)=4\sin^2(k_x/2)+4\sin^2(k_y/2)$ and $\omega_y(\bm k)=4\sin^2(k_y/2)$. The uniform mode decays at rate $\delta$. The $D_4\omega^2$ term suppresses short wavelengths, leaving a finite band in which gradients can grow. This finite-wavelength instability is analogous to that in the Swift–Hohenberg theory \cite{swift,CrossHohenberg1993}. Here the amplification coefficient depends on intermode coherence, so separability restricts the gain that can drive the instability.

For isotropic transport, or for modes with $k_y=0$ in the anisotropic case, continuous maximization over the Laplacian eigenvalue gives $\omega_*=(D_2e+\Dcl)/(2D_4)$. The maximizing value lies in the allowed band for the parameters used here. Thus
\begin{equation}
 \sigma_{\max}=-\delta+\frac{(D_2e+\Dcl)^2}{4D_4},\qquad
 e_c=\frac{2\sqrt{\delta D_4}-\Dcl}{D_2}.
 \label{eq:amplification-guide}
\end{equation}
Only discrete wavevectors are allowed on the finite lattice. With the same values of $r$, $\eta$, $\kappa_l$, and $\kappa_r$, all three parameter sets satisfy
\begin{equation}
 \boxed{2\ellinf<e_c<\einf.}\label{eq:causal-threshold}
\end{equation}
For the three parameter sets, $e_c\simeq0.61$--$0.78$ lies between $2\ellinf\simeq2\times10^{-4}$ and $\einf\simeq0.91$. The coherence permitted by separability cannot reach the instability threshold in this approximation. The coherence generated by the quantum source exceeds that threshold. Local entanglement enables the initial amplification, followed by nonlinear selection of the final morphology.

Section~\ref{sec:bounds} discusses the full noisy nonlinear evolution through bounds on pattern intensity for separable evolution, including spatial correlations and disorder. These bounds do not rely on the linear reduction.
\Needspace{14\baselineskip}
\section{Bounds for separable evolution}\label{sec:bounds}\label{sec:witness}
\subsection{Pattern intensity and admissible evolution}\label{sec:geometry-bound}\label{sec:comparison-class}
We quantify a target pattern by its band power, morphology and spatial extent. Let $g$ label the target pattern type and $\mathcal B$ a fixed band of nonzero wavevectors $\bm k$ in the reciprocal lattice. The Fourier amplitude and band power of an occupation configuration $u$ are
\begin{equation}
 A_{\bm k}(u)=\frac1M\sum_xe^{-i\bm k\cdot x}u_x,
 \qquad P_{\mathcal B}(u)=\sum_{\bm k\in\mathcal B}|A_{\bm k}(u)|^2,
 \qquad \PB(t)=\mathbb E[P_{\mathcal B}(u(t))].
 \label{eq:order}
\end{equation}
The diagonal amplitude operator $A_{\bm k}$ has eigenvalue $A_{\bm k}(u)$.
The mean squared contrast operator is $\widehat E=M^{-1}\sum_x\hat u_x^2$, with expectation $E(t)=\Tr[\rho(t)\widehat E]$. Parseval's identity gives $\PB(t)\le E(t)$.

Let $0\leq\chi_g(u)\leq1$ weight configurations according to the morphology and spatial extent of pattern $g$. The pattern intensity of a configuration $u$ is $G_g(u)=\chi_g(u)P_{\mathcal B}(u)$. We define $\widehat G_g$ to be diagonal in the physical number basis, with eigenvalue $G_g(u)$ for each occupation configuration, including configurations with leakage. The mean pattern intensity is
\begin{equation}
 \mathcal G_g(t)=\Tr[\rho(t)\widehat G_g]
 =\mathbb E[G_g(u(t))].
 \label{eq:Gdef}
\end{equation}
For the binary indicator used here, whenever $\Pr\{\chi_g=1\}>0$,
\begin{equation}
 \mathcal G_g(t)=\Pr\{\chi_g=1\}\,
 \mathbb E[P_{\mathcal B}\mid\chi_g=1].\label{eq:geometry-decomposition}
\end{equation}
The expectation includes all outcomes, with $G_g=0$ when the target criteria are not met.

The fixed indicator $\chi_g$ identifies both morphology and spatial extent. Stripes are distinguished by orientational order, spots by compact positive domains, and labyrinths by extended connected domains without a dominant orientation. Stripes and labyrinths must wind around a periodic direction, while spot centers must be distributed across the array. These criteria allow defects and multiple domains and are applied to both quantum and projected evolution.

Projection onto the physical number basis leaves the pattern intensity and the instantaneous occupation distribution unchanged. The comparison therefore concerns their dynamical generation and persistence.

We fix the partition into physical oscillator modes, the tensor product of the finite local spaces in Sec.~\ref{sec:local-space}, and the field, source, dephasing and leakage/repair channels in Eq.~\eqref{eq:gen}. Static coefficients may take any values in their specified ranges. The initial contrast and leakage conditions are specified separately for formation and persistence below.

An admissible control is nonselective and keeps the state within the specified occupation space. Between discrete interventions the unconditional density operator obeys $\dot\rho=(\Lind+\mathcal M_t)\rho$, where $\mathcal M_t$ is a time-dependent Lindblad generator. Define the mean leakage operator $\overline P_0=M^{-1}\sum_xP_{0,x}$. We require, for every density operator $\sigma$ on the specified support,
\[
 \Tr[\widehat E\,\mathcal M_t(\sigma)]\le0,
 \qquad \Tr[\overline P_0\,\mathcal M_t(\sigma)]=0.
\]
Thus control may decrease the second moment of the contrast but cannot add to it, and it leaves the leakage dynamics unchanged. A discrete intervention is a completely positive, trace-preserving map $\Phi$ satisfying the same support restriction and
\[
 \Phi^\dagger(\widehat E)\le\widehat E,
 \qquad \Phi^\dagger(\overline P_0)=\overline P_0
\]
on that support. These conditions refer to averages over all measurement outcomes, without postselection.

We consider admissible evolutions in which the unconditional density operator $\rho(t)$ remains separable across the individual physical modes at all times. Arbitrary classical correlations are allowed, and continuous-projection limits satisfying the same conditions are included. Projection onto the physical number basis preserves $\widehat E$ and $\overline P_0$. The continuous-time limit of repeated projection gives the projected generator derived in Appendix~\ref{app:projected-validation}. The bounds apply to all such evolutions, independently of the particular control used.

\subsection{Transport constraints and separable bounds}\label{sec:separable-bounds}
The local sum of squared differences is $f_x=\sum_{y\sim x}(u_x-u_y)^2$ on the square lattice. For a lattice state supported on the tensor product of the local spaces specified in Sec.~\ref{sec:local-space} and positive under partial transposition of mode $b_x$,
\begin{equation}
 \left|\langle C_xf_x\rangle\right|
 \le\left\langle P_{0,x}\left(2f_x+\frac{z\ee^2}{2}\right)\right\rangle,
 \qquad z=4.\label{eq:weighted}
\end{equation}
For positive symmetric weights, the same inequality holds with $f_x$ replaced by $f_{j,x}=\sum_yj_{xy}(u_x-u_y)^2$ and $z$ by $\sum_yj_{xy}$. The weighted inequality is derived in Appendix~\ref{app:weighted-derivation}.
Throughout this section, $\ell(t)=M^{-1}\sum_x\langle P_{0,x}\rangle_t$ is the mean leakage probability across the array. If the leakage probability is initially the same at every cell, this spatial average equals the single-cell value in Sec.~\ref{sec:coherence}. The transport contribution to the growth of the mean squared contrast is
\begin{equation}
 \mathcal T=\frac1M\sum_x\langle C_xf_{j,x}\rangle
 \le K_\ell\ell,\qquad K_\ell=32JU_e^2+2J\ee^2
 \quad\text{for separable states}.\label{eq:leakenergy}
\end{equation}
The uniform bound $K_\ell\ell$ remains valid when leakage and large local gradients are correlated. To retain these local leakage--gradient correlations, we define
\begin{equation}
 \mathcal W_{\mathrm{tr}}=
 \mathcal T-\frac{2J}{M}\sum_x\langle P_{0,x}f_x\rangle-2J\ee^2\ell
 \label{eq:transport-diagnostic}
\end{equation}
The quantity $\mathcal W_{\mathrm{tr}}$ is nonpositive for every state separable across the physical modes. A positive value in Fig.~\ref{fig:resource-geometry}(b) identifies transport beyond the local separable bound.

For the parameter regime $d>0$, defined in Appendix~\ref{app:bounds-derivation} and satisfied by all three parameter sets studied here, every admissible separable evolution obeys
\begin{equation}
 \dot E\le-\alpha E+c_0+c_1\ell(t),\label{eq:comparison}
\end{equation}
Here $\alpha>0$ describes net damping, $c_0$ is fixed by the rates and noise parameters, and $c_1\ell(t)$ accounts for leakage. Appendix~\ref{app:comparison-details} derives these constants. The long-time limit of the separable upper bound is
\begin{equation}
 B_\infty=\frac{c_0+c_1\ellinf}{\alpha}.\label{eq:comparison-floor}
\end{equation}

The operator $\hat G_g$ satisfies the inequality
\begin{equation}
 0\le\widehat G_g\le\sum_{\bm k\in\mathcal B}A_{\bm k}^\dagger A_{\bm k}\le\widehat E.
 \label{eq:geometry-order}
\end{equation}

First consider pattern formation from a weak initial field with $E(0)\le E_0$ and no initial leakage. Integrating Eq.~\eqref{eq:comparison} gives
\begin{align}
 0&\le\mathcal G_g(t)\le\PB(t)\le E(t)\le\Bsep(t),\label{eq:Gbound}\\
 \Bsep(t)&=E_0e^{-\alpha t}+B_\infty(1-e^{-\alpha t})
 -c_1\ellinf\frac{e^{-kt}-e^{-\alpha t}}{\alpha-k},\label{eq:mainbound}
\end{align}
where $k=\leak+\repair$.

The same comparison also bounds the persistence of an existing pattern. For a configuration at time $t_f$ with leakage fraction $\ell_f$, every admissible separable evolution starting from that configuration satisfies, at $\tau=t-t_f\ge0$,
\begin{align}
 \mathcal G_g(t_f+\tau)&\le\PB(t_f+\tau)\le B_f(\tau),\label{eq:maintenance-intensity}\\
 B_f(\tau)&=B_\infty+(U_e^2-B_\infty)e^{-\alpha\tau}
 +c_1(\ell_f-\ellinf)\frac{e^{-k\tau}-e^{-\alpha\tau}}{\alpha-k}.
 \label{eq:conditional-maint}
\end{align}
When $\alpha=k$, the quotients in Eqs.~\eqref{eq:mainbound} and \eqref{eq:conditional-maint} are replaced by their limits, $t e^{-\alpha t}$ and $\tau e^{-\alpha\tau}$, respectively. The bounds hold uniformly in the array size for all static coefficients in the specified ranges and for the nonselective controls defined in Sec.~\ref{sec:comparison-class}. The maintenance bound retains the leakage measured at $t_f$, including configurations with $\ell_f>\ellinf$. Exceeding the corresponding bound implies that the state was entangled at some time during the comparison interval. The local witness in Eq.~\eqref{eq:eln} separately tests the state at a specified time. Appendix~\ref{app:bounds-derivation} derives the differential inequality \eqref{eq:comparison} and the finite-time bounds \eqref{eq:Gbound}--\eqref{eq:conditional-maint}.

\subsection{Transport and the dynamical role of entanglement}\label{sec:resource-definition}
The excess pattern intensity, $\mathcal G_g(t)-\Bsep(t)$, constrains the transport required during formation. Define $\mathcal R_{\mathrm{tr}}=\mathcal T-K_\ell\ell$ by subtracting the uniform transport bound. In contrast, $\mathcal W_{\mathrm{tr}}$ uses local leakage--gradient correlations. Applying the preceding moment estimates without imposing separability on the transport term gives
\begin{equation}
 \boxed{\mathcal G_g(t)-\Bsep(t)
 \le\int_0^t e^{-\alpha(t-s)}\mathcal R_{\mathrm{tr}}(s)\,ds
 \le\int_0^t e^{-\alpha(t-s)}\mathcal W_{\mathrm{tr}}(s)\,ds.}
 \label{eq:geometry-transport-budget}
\end{equation}
An excess in pattern intensity yields a lower bound on the accumulated transport contribution beyond the separable limit. The exponential weight accounts for the decay of earlier contributions at rate $\alpha$. Equation~\eqref{eq:geometry-transport-budget} links the transport diagnostic in Fig.~\ref{fig:resource-geometry}(b) to the pattern intensity in Fig.~\ref{fig:resource-geometry}(c). Both sides of this integrated inequality have units of squared occupation contrast.

\Needspace{9\baselineskip}
We call formation from a weak initial field \emph{quantum self-organization} when the resulting mean pattern intensity exceeds the bound for all admissible separable evolutions. We call persistence under noise \emph{quantum stabilization} when evolution from the formed occupation configurations sustains a mean pattern intensity that exceeds its separable maintenance bound at the final observation time. The comparison fixes the partition into physical modes, the dynamical and control constraints, and the observation interval. These criteria concern the generation and persistence of spatial order, even when the occupation statistics at a given time can also be reproduced by a separable state.

\section{Pattern formation and stabilization}\label{sec:results}
\subsection{Formation beyond the separable bound}\label{sec:confirmation}
For all three parameter sets, evolution starts from the same weak Gaussian occupation field in a separable state with zero intermode coherence and no leakage. Varying the transport strength, nonlinear asymmetry and spatial anisotropy produces stripes, spots and labyrinths.

\begin{figure}[!htbp]
\centering\includegraphics[width=\linewidth]{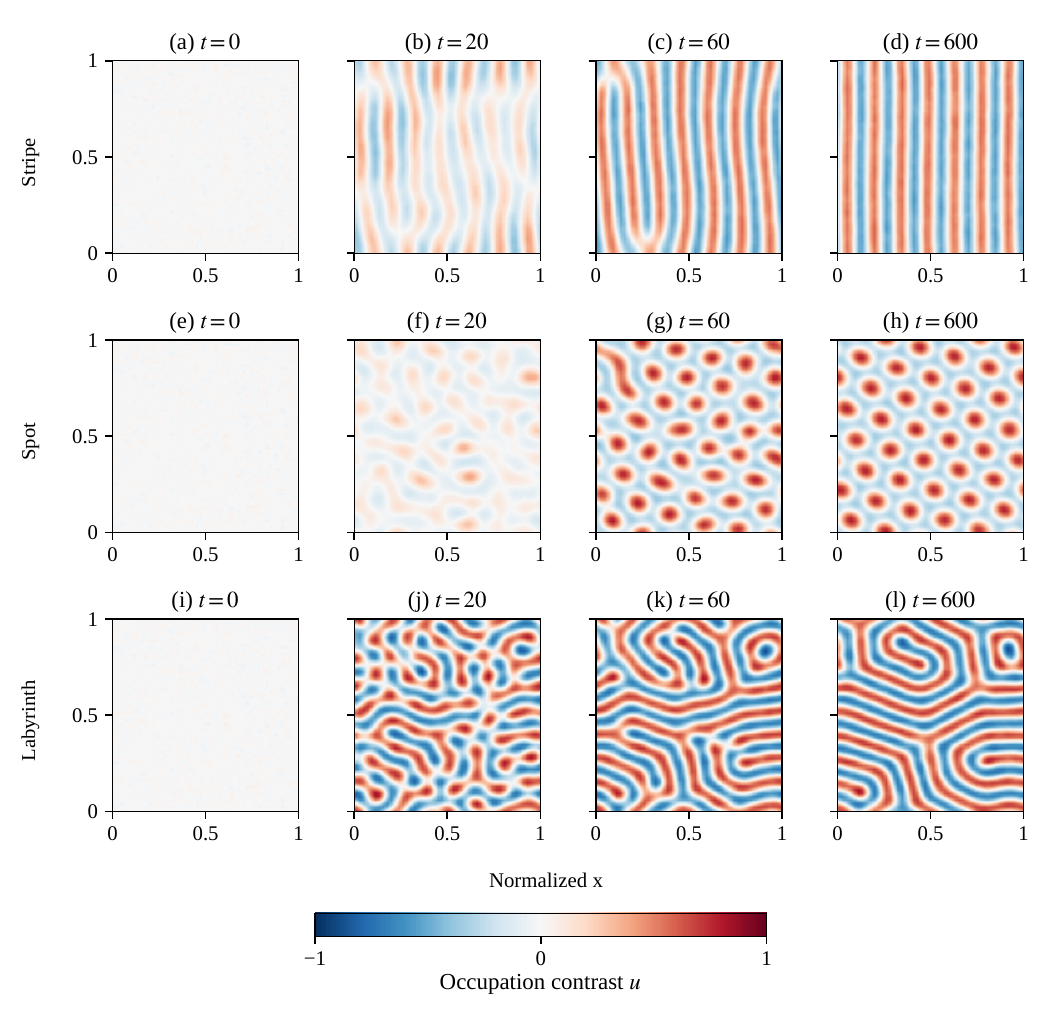}
\caption{Formation of stripe, spot and labyrinth patterns, shown in the top, middle and bottom rows. Each row follows one realization at $t=0,20,60,600$. All panels share the occupation-contrast scale and normalized spatial coordinates.}\label{fig:geometry-overview}
\end{figure}

The representative trajectories start from weak occupation fluctuations [Fig.~\ref{fig:geometry-overview}(a,e,i)], develop spatial modulation [Fig.~\ref{fig:geometry-overview}(b,f,j)], and form stripes, spots and labyrinths [Fig.~\ref{fig:geometry-overview}(c,g,k)]. These structures remain visible at the end of the formation interval [Fig.~\ref{fig:geometry-overview}(d,h,l)].

\begin{figure}[!htbp]
\centering\includegraphics[width=\linewidth]{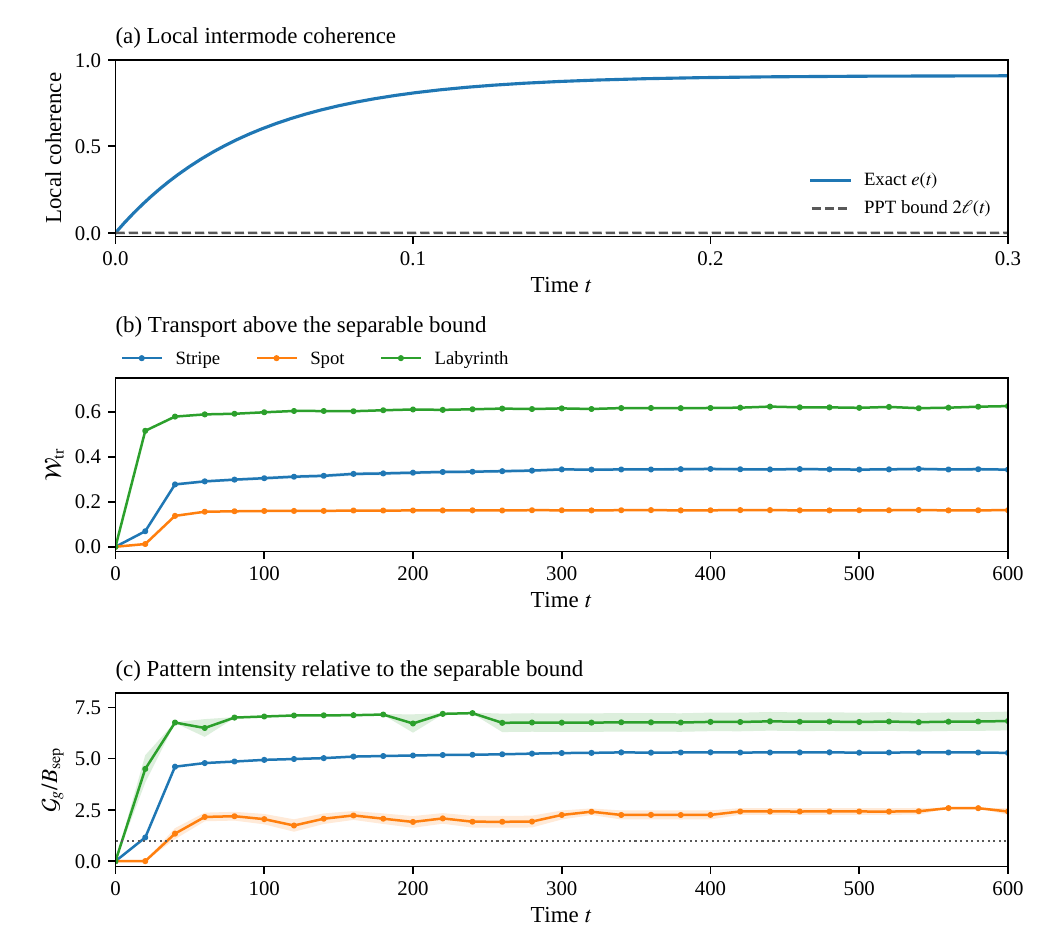}
\caption{From intermode coherence to extended patterns. (a) Exact local intermode coherence $e(t)$ and the separability bound $2\ell(t)$ during the initial local relaxation. (b) The transport diagnostic $\mathcal W_{\mathrm{tr}}$, positive when transport exceeds the local separable bound. (c) Mean pattern intensity divided by the formation bound. The horizontal dotted line at unity corresponds to the separable bound. Shading in (b,c) gives one standard error of the mean.}\label{fig:resource-geometry}
\end{figure}

Local intermode coherence develops before substantial spatial growth [Fig.~\ref{fig:resource-geometry}(a)]. As spatial gradients grow, $\mathcal W_{\mathrm{tr}}$ becomes positive, indicating transport beyond the separable bound [Fig.~\ref{fig:resource-geometry}(b)]. The simulated mean pattern intensities exceed the formation bounds by factors of about two to seven across the three parameter sets [Fig.~\ref{fig:resource-geometry}(c)]. Through Eq.~\eqref{eq:geometry-transport-budget}, an excess of the mean pattern intensity constrains the transport required to produce the patterns.

The mean pattern intensities in additional simulations with independently sampled initial fields and disorder also exceed the corresponding separable bounds, with greater variability in spot morphology.

\FloatBarrier
\subsection{Pattern persistence under noise}\label{sec:projected}
We compare the original quantum dynamics with evolution under continuous number-basis projection, starting from the same occupation and leakage configurations at $t_f$. Projection preserves the initial occupation distribution and keeps the subsequent state separable across the physical modes. We write $\tau=t-t_f$ for the elapsed time.

\begin{figure}[p]
\centering\includegraphics[width=\linewidth]{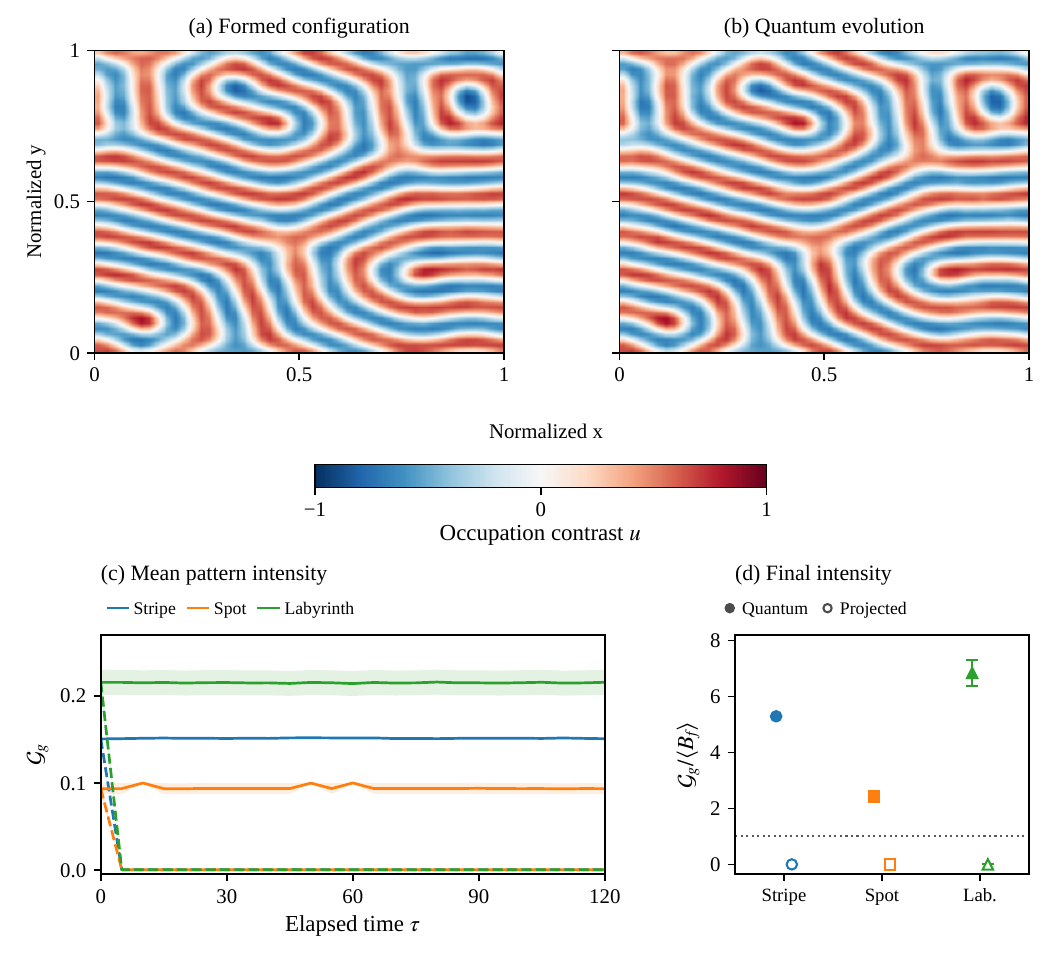}
\caption{Pattern persistence from the same initial occupation configurations. (a) A labyrinth at $t_f=600$ and (b) the same realization after quantum evolution to $\tau=120$, on a common contrast scale. (c) Mean pattern intensity under quantum (solid) and number-projected (dashed) evolution. Bands show one standard error of the mean. (d) Mean pattern intensity at $\tau=120$ divided by the mean maintenance bound for the starting configurations. Error bars give the standard error of the mean intensity divided by that mean bound. The horizontal line at unity corresponds to the separable bound.}\label{fig:geometry-maintenance}
\end{figure}

The number-basis projection is
\[
 \Delta(\rho)=\sum_{\boldsymbol n}|\boldsymbol n\rangle\langle\boldsymbol n|\rho|\boldsymbol n\rangle\langle\boldsymbol n|,
\]
where $\boldsymbol n$ labels all physical-mode occupations. On number-diagonal states, the projected evolution is generated by
\begin{equation}
 \Lind_{\mathrm{sep}}=\Delta\Lind\Delta.\label{eq:projected}
\end{equation}
This evolution satisfies the conditions for separable evolution specified in Sec.~\ref{sec:comparison-class}. Averaging the nonlinear jump rates over the local relative phase states fixes both its drift and its jump variance, as derived in Appendix~\ref{app:projected-validation}.

The labyrinth retains its extended bands and occupation contrast during continued quantum evolution [Fig.~\ref{fig:geometry-maintenance}(a,b)]. The mean pattern intensities remain nearly constant [Fig.~\ref{fig:geometry-maintenance}(c)] and exceed the separable maintenance bounds at the end of the continuation interval [Fig.~\ref{fig:geometry-maintenance}(d)]. Under continuous number-basis projection, the target pattern intensity becomes negligible during the subsequent evolution, while occupation fluctuations remain.
\FloatBarrier
\subsection{Suppression and recovery under dephasing}\label{sec:dephasing}
\begin{figure}[!htbp]
\centering\includegraphics[width=\linewidth]{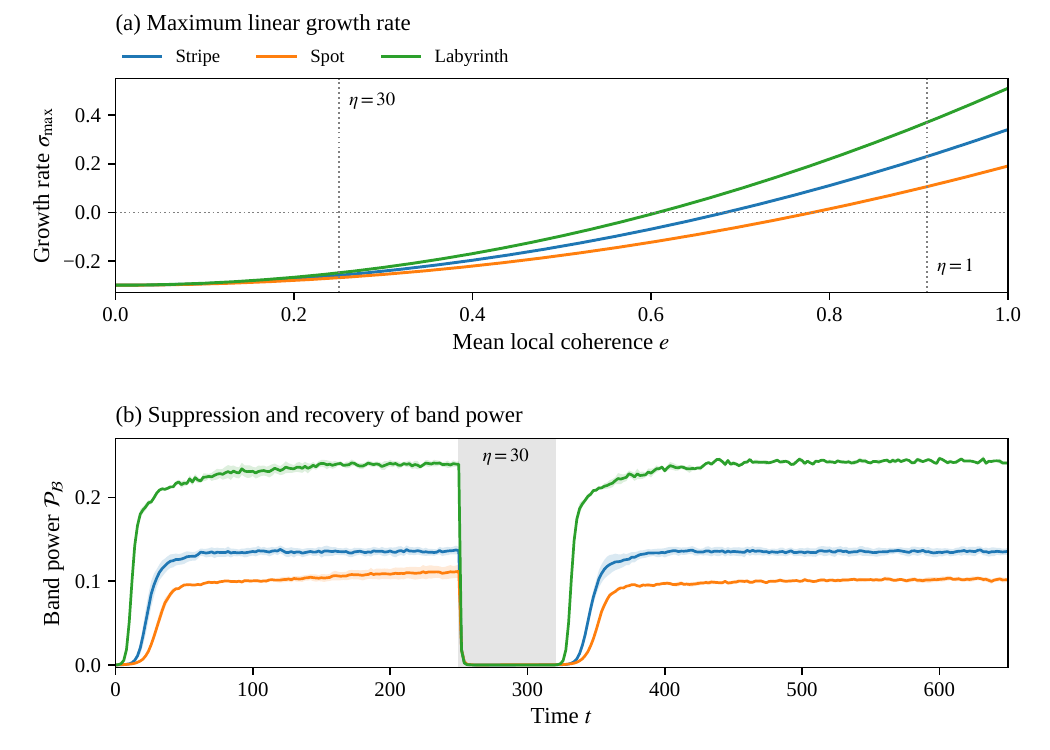}
\caption{Control of Turing amplification and recovery of band power. (a) Maximum linear growth rate near $u=0$ in the homogeneous deterministic limit with rapid phase relaxation. Vertical lines indicate stationary values of the local coherence at $\eta=1$ and $30$. (b) Mean band power under the dephasing cycle. Shading gives one standard error of the mean. The dephasing rate is $\eta=30$ in the gray interval $250\le t<320$ and $\eta=1$ otherwise. The band power returns close to its pre-dephasing level for all three parameter sets.}\label{fig:source-control}
\end{figure}

Continuous projection onto the number basis keeps the state separable throughout the evolution. Varying the dephasing rate controls the coherence available for amplification. We vary $\eta:1\to30\to1$ while holding all other field rates and noise parameters fixed. In the linear analysis, increasing the dephasing rate lowers the stationary coherence and changes the maximum growth rate from positive to negative for all three parameter sets [Fig.~\ref{fig:source-control}(a)]. At $\eta=30$, the stationary coherence is approximately $0.25$, below the thresholds in Eq.~\eqref{eq:amplification-guide}. The band power decreases during this interval of strong dephasing and recovers after the original dephasing rate is restored [Fig.~\ref{fig:source-control}(b)].

The stationary values of $e$ and $\ell$ at $\eta=30$ still satisfy $e_\infty>2\ell_\infty$, so local entanglement remains. The suppression and recovery of band power therefore reflect changes in the coherence-dependent transport gain. 

\FloatBarrier

\section{Autonomous classicalization}\label{sec:Classicalization}
We couple each bosonic dimer to a two-state environmental degree of freedom that responds to the developing occupation contrast. Environmental transitions remove local intermode coherence, suppress its regeneration and activate classical transport. This allows order formed through quantum dynamics to persist as local entanglement is lost. All model parameters are time independent.

Let $h_x=0,1$ label the environmental states and $H_x=|1\rangle\langle1|_{\mathrm{env},x}$ the projector onto $h_x=1$. In the field drift of Eq.~\eqref{eq:F}, replace $C_x$ by
\begin{equation}
 Q_x=(1-H_x)C_x+\chi H_x,
 \label{eq:classicalization_transport}
\end{equation}
and replace the coherence-generating jump $\sqrt{r}\,R_x$ by $\sqrt{r}\,(1-H_x)R_x$. The number shifts, dephasing, leakage, repair and background occupation noise retain their original form. For $h_x=0$, the contribution to transport depends on intermode coherence. For $h_x=1$, it is set by the constant $\chi$, and the local coherence source is inactive. The coefficient $\Dcl$ remains unchanged.

The irreversible environmental transition $h_x=0\to1$ depends on the squared occupation contrast averaged over the cell and its four nearest neighbors,
\begin{align}
 A_x
 &=\frac{u_x^2+\sum_{y\sim x}u_y^2}{5},
 \label{eq:classicalization_activity}\\
 \Lambda_x
 &=\Gamma\frac{A_x^4}{A_x^4+A_*^4}
   +\frac{\beta}{4}\sum_{y\sim x}H_y.
 \label{eq:classicalization_rate}
\end{align}
The first term describes a local response with rate scale $\Gamma$, which becomes appreciable when $A_x$ is comparable to $A_*$. The second term, controlled by $\beta$, allows transitions in neighboring cells to promote the same change at $x$, including in regions of small contrast. For each physical number state $|n,\sigma\rangle_x$, with $\sigma=L,R,0$, we add the jump operator
\begin{equation}
 K_{x,n,\sigma}
 =|1\rangle\langle0|_{\mathrm{env},x}
  |n,\sigma\rangle_x\langle n,\sigma|\sqrt{\Lambda_x}.
 \label{eq:classicalization_jump}
\end{equation}
Adding $\sum_{x,n,\sigma}\D[K_{x,n,\sigma}]$ to the modified generator completes the model. The quantities $A_x$ and $\Lambda_x$ are diagonal operators constructed from the occupations and environmental projectors. The jump operators are therefore time independent, and the master equation remains linear in the density operator. Each environmental jump projects the local dimer onto a physical number state without transferring particles. Once $h_x=1$, the remaining channels preserve diagonality in the physical number basis.

Initially, the occupation field has weak fluctuations,
the dimers have no intermode coherence or leakage, and all
environmental cells have $h_x=0$. The local source generates
coherence before the environment changes appreciably.
The parameter values and numerical methods are given in
Appendix~\ref{app:methods}. When all environmental cells have
reached state $1$, the amplification coefficient in the
homogeneous limit is $\Dcl+\chi D_2$. For the parameters used
here, this exceeds the instability threshold
$2\sqrt{\delta D_4}$ of Eq.~\eqref{eq:amplification-guide}.

\begin{figure}[H]
 \centering
 \includegraphics[
   width=\linewidth,
   height=0.82\textheight,
   keepaspectratio
 ]{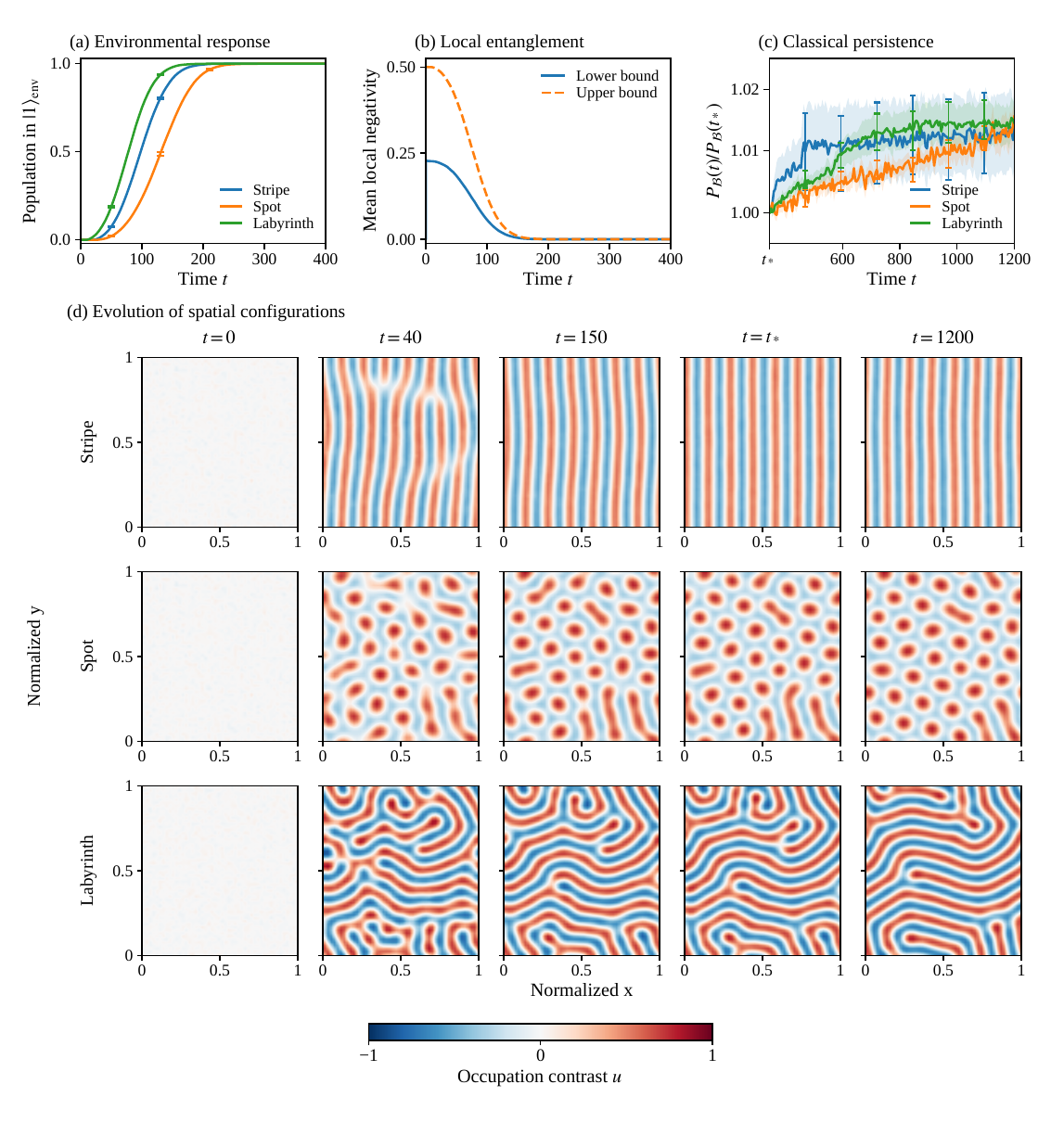}
 \caption{
 Autonomous classicalization and persistence of spatial order. (a) Mean fraction of environmental cells in state $h_x=1$. (b) Estimates of the lower and upper bounds on the mean local negativity in Eq.~\eqref{eq:classicalization_negativity} for the labyrinth ensemble. (c) Mean of the individual ratios of band power $P_{\mathcal B}(t)/P_{\mathcal B}(t_*)$. The reference time $t_*=345$ is the first observation time at which every environmental cell has reached state $1$ in all sampled realizations. Error bars in (a,c) and shading in (c) indicate one standard error over sixteen realizations per parameter set. (d) Representative configurations on a common occupation-contrast scale. Each row follows one realization, selected near the median final band power for its parameter set. The Fourier band is fixed throughout the evolution.
 }
 \label{fig:classicalization}
\end{figure}

For the initial states considered here, the state remains separable across lattice cells, although the two modes within each cell may be entangled. The local states are diagonal in the $|n,+\rangle$, $|n,-\rangle$ and $|n,0\rangle$ basis for $h_x=0$, and in the physical number basis for $h_x=1$. The populations therefore evolve according to a continuous-time Markov process. Figure~\ref{fig:classicalization}(a) shows the fraction of environmental cells that have reached state $1$ as spatial order develops.

For the unconditional two-mode reduced states $\rho_x$, the mean local negativity obeys
\begin{equation}
 \frac{
   \left[M^{-1}\sum_x\langle C_x-2P_{0,x}\rangle\right]_+
 }{4}
 \leq
 \frac{1}{M}\sum_x\mathcal N(\rho_x)
 \leq
 \frac{1}{2M}\sum_x
 \left\langle(1-H_x)(1-P_{0,x})\right\rangle.
 \label{eq:classicalization_negativity}
\end{equation}
The lower bound follows from Eq.~\eqref{eq:eln}. Convexity of the negativity gives the upper bound because each local basis state $|n,\pm\rangle$ has negativity $1/2$, whereas the leakage and physical number states are separable. Figure~\ref{fig:classicalization}(b) shows estimates of both bounds for the labyrinth ensemble. The upper bound is one half of the mean fraction of cells with $h_x=0$ that are outside the leakage sector. Conditioned on every environmental cell having reached state $1$, the physical state is diagonal in the number basis and remains separable. Contributions from environmental configurations with some $h_x=0$ are retained in the unconditional upper bound.

Let $t_*$ denote the first observation time at which every environmental cell has reached state $1$ in all simulated realizations. The Lindblad generator remains unchanged at $t_*$. Figure~\ref{fig:classicalization}(c) shows the mean of $P_{\mathcal B}(u(t))/P_{\mathcal B}(u(t_*))$ over all realizations, using the same fixed Fourier band throughout the evolution. The mean ratio remains close to unity during the subsequent classical evolution, and most realizations retain the target morphology. Figure~\ref{fig:classicalization}(d) shows representative configurations through the formation of spatial order, the environmental response and subsequent classical evolution.

Additional simulations isolate the contributions of the two transport mechanisms. With the same initial conditions and environmental response, removing both the coherence source and the coherence-dependent transport for $h_x=0$ produces neither environmental transitions nor target patterns during the simulated interval. Calculations with the exact number-projected dynamics give the same outcome. Keeping the environmental transitions but setting $\chi=0$ allows initial pattern formation, followed by decay to the noise level. When the environmental transitions are disabled, the coherence source remains active and the patterns persist. The same sequence of autonomous classicalization and persistent order also occurs in simulations with nonzero initial intermode coherence.

The separable bounds in Sec.~\ref{sec:bounds} apply to the fixed channels of Eq.~\eqref{eq:gen}. In the present extension, the environment activates an additional contribution to transport that can maintain order without intermode coherence. The resulting dynamics connects quantum formation and classical maintenance within a single autonomous open system.

\section{Discussion}
We establish entanglement as a necessary resource for the pattern intensities attained in the bosonic lattice under
the stated dynamical constraints. Intermode coherence controls the transport that amplifies spatial variations,
while separability limits the intensity that the full nonlinear dynamics can generate or sustain. Quantum
evolution exceeds these limits during both formation and persistence under noise [Eqs.~\eqref{eq:Gbound}--\eqref{eq:conditional-maint}, Figs.~\ref{fig:resource-geometry}(c) and~\ref{fig:geometry-maintenance}(c,d)].
Entanglement remains confined to individual dimers, while finite-range transport connects its local effects to
order across the lattice.

The distinction between occupation statistics and the dynamics that generate them is central to these results.
A spatial pattern alone does not determine whether entanglement was needed to produce or sustain it.
Number-basis projection preserves the initial occupation distribution but changes the subsequent evolution
[Eq.~\eqref{eq:projected}, Fig.~\ref{fig:geometry-maintenance}(c,d)]. Under the specified constraints, the separable bounds allow occupation measurements to establish the need for entanglement during the evolution.
Equation~\eqref{eq:geometry-transport-budget} further relates the excess intensity to transport beyond the
separable limit. Macroscopic observables can therefore reveal the dynamical role of a local quantum resource
without reconstructing the full many-body state.

The model with environmental degrees of freedom shows how the resulting order can persist after local entanglement is lost. As occupation variations grow, they induce environmental transitions that remove local coherence and activate classical amplification [Eqs.~\eqref{eq:classicalization_transport} and~\eqref{eq:classicalization_jump}]. As the environmental transitions proceed, the upper bound on local entanglement approaches zero while the patterns persist [Eq.~\eqref{eq:classicalization_negativity}, Fig.~\ref{fig:classicalization}(b--d)]. This sequence follows from a time-independent generator, without prescribing a switching time. The resources required to form order can therefore differ from those sufficient to maintain it within a single autonomous open system.

The dephasing and environmental extensions reveal two complementary regimes. Strong dephasing suppresses spatial amplification while local entanglement persists [Sec.~\ref{sec:dephasing}, Fig.~\ref{fig:source-control}]. In the environmental extension, spatial order survives as local entanglement is lost. The stability of macroscopic order therefore depends on the transport mechanism that sustains it.

These findings suggest a broader approach to nonequilibrium emergence based on the resources required to form collective order and those needed to sustain it. An important next question is how the amount, spatial distribution and lifetime of entanglement determine the formation, stability and recovery of macroscopic structures. Extending the present bounds to transport, synchronization and other collective phenomena could establish when transient quantum resources are sufficient to generate order that persists under classical dynamics. This approach may provide a basis for harnessing quantum effects in autonomous nonequilibrium systems.

\subsection*{Data and code availability}
The data and code supporting this study are provided in the author's GitHub repository \url{https://github.com/IKEDAKAZUKI/Quantum-self-organization-and-stabilization}

\subsection*{Acknowledgment}
This work was supported by the NSF under Grant No. OSI-2328774. 

\FloatBarrier
\appendix
\section{Local channels and coherence dynamics}\label{app:model-details}\label{app:source-transient}
The leakage and repair operators in the generator \eqref{eq:gen} are
\begin{align}
 V_{x,L}&=\sqrt{\leak}\sum_n|n,0\rangle\langle n,L|,&
 V_{x,R}&=\sqrt{\leak}\sum_n|n,0\rangle\langle n,R|,\\
 W_{x,L}&=\sqrt{\repair/2}\sum_n|n,L\rangle\langle n,0|,&
 W_{x,R}&=\sqrt{\repair/2}\sum_n|n,R\rangle\langle n,0|. \label{eq:leakjumps}
\end{align}
The sums run over $0\le n\le n_{\max}$. Leakage removes the extra particle from either imbalanced configuration, and repair restores it to $L$ or $R$ with equal probabilities. Their total rates are $\leak$ and $\repair$.

For $\ell(0)=0$ and initial coherence $e(0)$, Eq.~\eqref{eq:source} gives
\begin{equation}
 \ell(t)=\ellinf(1-e^{-kt}),\qquad k=\leak+\repair.
 \label{eq:leakage-transient}
\end{equation}
\begin{equation}
 e(t)=e(0)e^{-at}+\einf(1-e^{-at})
 +r\ellinf\frac{e^{-kt}-e^{-at}}{a-k},
 \qquad a=r+2\eta+\leak .\label{eq:exact-source}
\end{equation}
At $a=k$, the quotient is understood by continuity as $t e^{-at}$.

\section{Derivation of the separable bounds}\label{app:comparison-details}
\subsection{Weighted partial-transpose inequality}\label{app:weighted-derivation}
We derive the local entanglement bounds \eqref{eq:eln} and the weighted transport inequality \eqref{eq:weighted}.
The state $|n_{\max}+1,n_{\max}+1\rangle$ lies outside the invariant subspace and has zero population. Its projector is included in $P_0$ so that the boundary contribution remains nonnegative under partial transposition. For the local witness $2P_0-C$, the partial transpose is positive on the chain of equal-occupation states $\{|n,n\rangle:0\le n\le n_{\max}+1\}$ and has operator norm at most four. This gives the negativity bound in Eq.~\eqref{eq:eln}.

To derive Eq.~\eqref{eq:weighted}, consider a lattice state supported on the tensor product of the local spaces in Sec.~\ref{sec:local-space} and positive under partial transposition of mode $b_x$. For positive symmetric weights $j_{xy}$, the same derivation replaces $f_x$ by $f_{j,x}=\sum_yj_{xy}(u_x-u_y)^2$ and $z$ by $\sum_yj_{xy}$ in the correction term.

The partial transpose of $C_x f_x$ has off-diagonal matrix elements between $|n,n\rangle$ and $|n+1,n+1\rangle$, weighted by $f_x$ evaluated midway between the occupation contrasts of the two states. With all neighboring occupations held fixed, shifting the contrast at cell $x$ by $\pm\ee/2$ gives
\begin{equation}
 f_x(u+\ee/2)+f_x(u-\ee/2)=2f_x(u)+z\ee^2/2.
\end{equation}
For either sign, the partial transpose of $P_0(2f_x+z\ee^2/2)\pm C_xf_x$ is unitarily equivalent to a weighted Laplacian on the chain of equal-occupation states, with additional nonnegative diagonal terms at the boundaries. For the sign with positive off-diagonal entries, the unitary changes the sign of alternating equal-occupation basis vectors. This holds for each fixed configuration of all remaining occupations. Taking the expectation in a PPT state and summing over the remaining occupations gives the inequality for both signs. Positive edge weights give the stated extension. Summing the weighted inequality over cells gives the transport bound \eqref{eq:leakenergy}. Bounding each edge weight by $J$ while keeping the local leakage--gradient correlations gives $\mathcal W_{\mathrm{tr}}\leq0$ (Eq.~\eqref{eq:transport-diagnostic}) for states separable across the physical modes.

\subsection{Derivation of the formation and maintenance bounds}\label{app:bounds-derivation}
We derive the differential inequality \eqref{eq:comparison} from the conditional moments \eqref{eq:moments}, then integrate it to obtain the formation and maintenance bounds \eqref{eq:Gbound}--\eqref{eq:conditional-maint}. The quantity being bounded is the mean squared contrast, $E=M^{-1}\sum_x\langle u_x^2\rangle$.

The following coefficients separate the damping of $E$
from the terms that can increase it. Here $d$ denotes the net damping remaining after bounding the local nonlinear and coherence-independent amplification terms. The constant $C_{\rm nl}$ comes from the nonlinear terms, and $C_{\rm diff}$ includes it in the bound on the variance of the field jumps. The quantity $g_\ee$ bounds the increase of $u_x^2$ in one leakage or repair event. Choose $0<\theta<1$ with $d>0$. Set
\begin{align}
 d&=2\delta_- -16\Dcl-\frac{\nu^2}{2b(1-\theta)}-\ee|\nu|>0,
 &\alpha&=d/2, \label{eq:d}\\
 C_{\rm nl}&=\frac{27b\ee^4}{2048\theta^3}+\frac{5^5c\,\ee^6}{2^5 6^6},\label{eq:cnl}\\
 C_{\rm diff}&=\ee^2\left[\frac{\delta_+^2}{d}
 +\frac{(8D_4+\Dcl)^2}{4D_4}+\frac\zeta2
 +\frac{2J^2}{\sqrt{dD_4}}\right]+C_{\rm nl}, \label{eq:cdiff}\\
 g_\ee&=\ee U_e+\ee^2/4,
 &c_0&=C_{\rm diff}+\ee s_0+2\Db+g_\ee\leak,\label{eq:gleak}\\
 c_1&=32JU_e^2+2J\ee^2+g_\ee\repair,
 &B_\infty&=(c_0+c_1\ellinf)/\alpha. \label{eq:constants}
\end{align}

The field jumps contribute to $\dot E$ through the conditional drift and second moment in Eq.~\eqref{eq:moments}. Hence
\[
 \left.\frac{dE}{dt}\right|_{\mathrm{field}}
 \le\frac1M\sum_x\left\langle
 2u_xF_x+\ee\sqrt{F_x^2+s_0^2}+2\Db\right\rangle
 \le\frac1M\sum_x\langle2u_xF_x+\ee|F_x|\rangle+\ee s_0+2\Db,
\]
where the second inequality uses $\sqrt{F_x^2+s_0^2}\le |F_x|+s_0$. The first relation is an equality away from the occupation cutoffs. At a cutoff, the omitted outward jumps would increase $u_x^2$, so the upper estimate remains valid.

The nonlinear part of the drift \eqref{eq:F} contributes $2\nu u_x^3-2bu_x^4-2cu_x^6$ to $2u_xF_x$.
The deterministic local terms obey
\begin{equation}
 2\nu u^3\le2b(1-\theta)u^4+\frac{\nu^2}{2b(1-\theta)}u^2.\label{eq:quadratic-drift-bound}
\end{equation}
Bounding $\ee|F_x|$ using Eq.~\eqref{eq:F} adds the nonlinear terms $\ee|\nu|u_x^2$, $\ee b|u_x|^3$ and $\ee c|u_x|^5$. After Eq.~\eqref{eq:quadratic-drift-bound}, the remaining nonlinear damping controls the cubic and quintic terms through $\ee b|u|^3-2b\theta u^4\le27b\ee^4/(2048\theta^3)$ and $\ee c|u|^5-2cu^6\le 5^5c\,\ee^6/(2^5 6^6)$. These constants form $C_{\rm nl}$ in Eq.~\eqref{eq:cnl}. The term $\ee|\nu|u^2$ is included in the definition of $d$.

For the spatial terms, define $A=M^{-1}\langle u^TL^2u\rangle$, $B=M^{-1}\langle u^TL_yu\rangle$ and $X=M^{-1}\langle u^TLu\rangle$. They satisfy $X\le8E$ and $X\le\sqrt{EA}$. The deterministic spatial damping contributes $-2D_4A-2\zeta B$. The coherence-independent amplification is bounded by $2\Dcl X\le16\Dcl E$, and the weighted inequality in Appendix~\ref{app:weighted-derivation} bounds the remaining transport by $K_\ell\ell$.

We retain the negative gradient terms when bounding the contributions to the jump variance from the part of $F_x$ linear in $u$. These contributions are at most
\begin{equation}
 \ee\delta_+\sqrt E+\ee(8D_4+\Dcl)\sqrt A
 +2\ee\zeta\sqrt B+\ee J\sqrt{8X}.
\end{equation}
The first three terms are absorbed by $dE/4$, $D_4A$ and $2\zeta B$, respectively. The last term is absorbed by the remaining $dE/4+D_4A$ using $X\le\sqrt{EA}$. Completing the square yields the coefficients in Eq.~\eqref{eq:cdiff}, leaving $-dE/2$. Leakage and repair change $u$ by $\mp\ee/2$ and contribute at most $g_\ee(\leak+\repair\ell)$. The additional control in Sec.~\ref{sec:comparison-class} contributes $\Tr[\widehat E\,\mathcal M_t(\rho)]\le0$ and leaves $\dot\ell=\leak(1-\ell)-\repair\ell$ unchanged. Therefore
\begin{equation}
 \dot E\le-\alpha E+c_0+c_1\ell(t). \label{eq:comparison-detail}
\end{equation}
At a discrete admissible intervention, $E$ cannot increase and $\ell$ is unchanged. The same scalar comparison therefore holds across such interventions and in their continuous-projection limit. For formation, integrating Eq.~\eqref{eq:comparison-detail} with $E(0)\le E_0$ and the leakage profile \eqref{eq:leakage-transient} gives Eq.~\eqref{eq:mainbound}. The operator inequality \eqref{eq:geometry-order} then gives Eq.~\eqref{eq:Gbound}.

For persistence from a configuration at $t_f$, we use $E(t_f)\le U_e^2$ and the conditional leakage profile $\ell(\tau)=\ellinf+(\ell_f-\ellinf)e^{-k\tau}$. The same integration gives Eq.~\eqref{eq:conditional-maint}, including its configuration-dependent leakage correction. The operator inequality \eqref{eq:geometry-order} yields Eq.~\eqref{eq:maintenance-intensity}. The inequalities involving occupation numbers hold at the operator level, with the operators diagonal in the physical number basis. Separability is required only for the weighted transport bound.

\subsection{Bound on the integrated transport contribution}\label{app:integrated-transport}

We derive Eq.~\eqref{eq:geometry-transport-budget} by retaining the transport contribution $\mathcal T$ in the estimates of Appendix~\ref{app:bounds-derivation}. For the dynamics, initial conditions and controls specified in Sec.~~\ref{sec:comparison-class}, these estimates give
\begin{equation}
 \dot E\le-\alpha E+c_0+g_\ee\repair\ell(t)+\mathcal T(t)
 =-\alpha E+c_0+c_1\ell(t)+\mathcal R_{\mathrm{tr}}(t),
 \qquad \mathcal R_{\mathrm{tr}}=\mathcal T-K_\ell\ell.
 \label{eq:unrestricted-budget}
\end{equation}
The other estimates use positivity, operator bounds and dissipative terms and do not require separability. Subtracting the differential equation satisfied by $\Bsep$ gives
\[
 \frac{d}{dt}(E-\Bsep)\le-\alpha(E-\Bsep)+\mathcal R_{\mathrm{tr}},
 \qquad E(0)-\Bsep(0)\le0.
\]
Multiplying by the integrating factor $e^{\alpha t}$ and integrating gives an upper bound on $E(t)-\Bsep(t)$ in terms of the exponentially weighted integral of $\mathcal R_{\mathrm{tr}}$. Using $\mathcal G_g\le E$ gives the first inequality in Eq.~\eqref{eq:geometry-transport-budget}. For the second inequality, $f_x\le16U_e^2$ on the square grid implies
\begin{equation}
 2J M^{-1}\sum_x\langle P_{0,x}f_x\rangle+2J\ee^2\ell
 \le K_\ell\ell,
\end{equation}
so $\mathcal R_{\mathrm{tr}}\le\mathcal W_{\mathrm{tr}}$. All expectations in Eq.~\eqref{eq:geometry-transport-budget} are evaluated with the
unconditional density operator. Figure~\ref{fig:resource-geometry}(b) plots $\mathcal W_{\mathrm{tr}}$, which enters the integrated bound \eqref{eq:geometry-transport-budget}.

\Needspace{10\baselineskip}
\section{Dynamics under number-basis projection}\label{app:projected-validation}

For the field jumps in Eq.~\eqref{eq:fieldjumps}, projection onto the physical number basis averages the transition rates over the local
eigenvalues of $C_x$.

A physical number state in the nonleaked sector has equal weights in the $C=+1$ and $C=-1$ sectors. The projected field rate is obtained by averaging over these signs, while leaked cells have $C=0$. Write $F_x=F_x^{(0)}+\sum_{y\sim x}j_{xy}(C_x+C_y)(u_x-u_y)/2$, where $F_x^{(0)}$ contains every term independent of $C$. For fixed occupations and leakage configuration, assign an independent sign $\sigma_z=\pm1$ to each nonleaked cell in $\{x\}\cup\{y:y\sim x\}$ and set $\sigma_z=0$ on leaked cells. Let $m_x\le5$ be the number of nonleaked cells in this neighborhood. The total projected rates, including the background occupation
jumps, are
\begin{equation}
 \overline q_x^{\,\pm}
 =2^{-m_x}\sum_{\boldsymbol\sigma}
 \frac{\sqrt{F_x(\boldsymbol\sigma)^2+s_0^2}\pm F_x(\boldsymbol\sigma)}{2\ee}
 +\frac{\Db}{\ee^2}.
 \label{eq:projectedrates}
\end{equation}
The same occupation cutoffs apply as in the quantum generator. Away from the occupation cutoffs, the projected field jumps contribute the conditional drift and variance
\begin{align}
 \mathbb E[du_x\mid u,\mathrm{leak}]/dt&=F_x^{(0)},\\
 \mathbb E[(du_x)^2\mid u,\mathrm{leak}]/dt
 &=\ee\,2^{-m_x}\sum_{\boldsymbol\sigma}\sqrt{F_x(\boldsymbol\sigma)^2+s_0^2}+2\Db.
 \label{eq:projectedvariance}
\end{align}
The sign average in Eq.~\eqref{eq:projectedrates} determines both the drift and the variance in Eq.~\eqref{eq:projectedvariance}. The variance retains the nonlinear square-root average and generally differs from the result obtained by setting $C=0$ in the original rate.

To obtain Eq.~\eqref{eq:projectedrates}, expand each number state $|n,L\rangle$ or $|n,R\rangle$ in the relative-phase basis. The two basis components have equal probabilities. Summing the squared transition amplitudes over the final $L/R$ occupation states gives the sign average in Eq.~\eqref{eq:projectedrates}, independently of the initial $L/R$ state. Leakage and repair retain rates $\leak$ and $\repair$. Summing over the $L/R$ occupation labels gives a closed Markov process for the occupation and leakage variables. Transitions between $L$ and $R$ at fixed total occupation do not affect this reduced process.

Projection onto the physical number basis satisfies $\Delta^\dagger(\widehat E)=\widehat E$ and $\Delta^\dagger(\overline P_0)=\overline P_0$. For states diagonal in the physical number basis, $\Delta e^{h\Lind}\Delta=\Delta+h\Delta\Lind\Delta+O(h^2)$. Since the occupation space is finite, repeated projection has the limit
\[
 \lim_{n\to\infty}(\Delta e^{t\Lind/n}\Delta)^n\rho
 =e^{t\Delta\Lind\Delta}\Delta\rho.
\]
The limiting state is diagonal in the physical number basis, hence separable at every time. For $E$ and $\ell$, the projected and original generators give the same time derivatives when evaluated on this diagonal state. The evolution therefore satisfies the differential inequality derived in Appendix~\ref{app:bounds-derivation}. This continuous-projection limit is the specific separable control used in the simulations.

\section{Numerical methods}\label{app:methods}\label{sec:numerical-settings}
The bosonic parameters are listed in Table~\ref{tab:model-parameters}. For the environmental model in Sec.~\ref{sec:Classicalization}, we retain these parameters and set
\begin{equation}
 \Gamma=0.002,\qquad
 \beta=0.08,\qquad
 A_*=0.10,\qquad
 \chi=0.91.
 \label{eq:classicalization_parameters}
\end{equation}
Times are measured in units of $D_4^{-1}$, with $D_4=1$, and the separable bounds are evaluated with $\theta=0.15$. Pattern formation starts from a state diagonal in the physical number basis, with Gaussian occupation fluctuations of root-mean-square contrast $0.02$, zero intermode coherence and no leakage. The environmental cells in Sec.~\ref{sec:Classicalization} initially have $h_x=0$. The initial second moment $E_0=M^{-1}\sum_xu_x(0)^2$ determines the formation bound in Eq.~\eqref{eq:mainbound}.

\begin{table}[htbp]\centering\small
\begin{tabular}{@{}lrlr@{}}\toprule
\multicolumn{4}{c}{Common parameters}\\\midrule
Parameter & Value & Parameter & Value \\\midrule
$\delta$ & 0.3 & $D_4$ & 1 \\
$b$ & 1 & $c$ & 0.1 \\
$r$ & 20 & $\eta$ & 1 \\
$\leak$ & $2\times10^{-4}$ & $\repair$ & 2 \\
$\Db$ & $10^{-5}$ & $s_0$ & 0.002 \\
$\Dcl$ & 0.001 & $w$ & 0.05 \\
$U$ & 1 & $\ee$ & 0.01 \\\midrule
Parameter set & $D_2$ & $\nu$ & $\zeta$ \\\midrule
Stripe & 1.6 & 0 & 0.8 \\
Spot & 1.4 & 0.55 & 0 \\
Labyrinth & 1.8 & 0 & 0 \\\bottomrule
\end{tabular}
\caption{Bosonic parameters for the formation, persistence and autonomous classicalization simulations. The three sets differ only in transport strength $D_2$, nonlinear asymmetry $\nu$ and spatial anisotropy $\zeta$.}\label{tab:model-parameters}
\end{table}

We simulate the continuous-time population dynamics using the Gillespie algorithm~\cite{gillespie}. The original bosonic model is sampled in the joint $|n,s\rangle$ basis. For Sec.~\ref{sec:Classicalization}, the sampled states also include the environmental variables and the corresponding local basis specified there.

The formation, persistence and autonomous classicalization simulations use $48\times56$ cells with periodic boundaries and unit lattice spacing. For the formation and persistence comparisons in Sec.~\ref{sec:results}, each parameter set has sixteen independent stochastic realizations with the initial field and disorder held fixed. Quantum and number-projected evolutions start from the same occupation and leakage configurations. In Sec.~\ref{sec:Classicalization}, each parameter set has sixteen realizations with independently sampled initial fields and disorder. The dephasing cycle uses $24\times28$ cells and four realizations per parameter set.

Additional formation and persistence calculations use four independent initial-field and disorder pairs, with two stochastic realizations for each pair. Standard errors for these calculations are evaluated from the four pair-averaged intensities.

The spectral band comprises the unstable modes of Eq.~\eqref{eq:linear-guide} at $e=1$ on the finite reciprocal lattice. The same fixed criteria for morphology and spatial extent apply to quantum and projected evolution. Shaded bands and error bars indicate one standard error of the mean for the numerical averages.

For each parameter set, we test the separable bounds at the end of formation, $t_f=600$, and after quantum evolution for $\tau=120$. For realization $i$, let $G_i=G_g(u_i)$ be the final pattern intensity and let $B_i$ be the corresponding bound from Eq.~\eqref{eq:mainbound} or Eq.~\eqref{eq:conditional-maint}, evaluated with that realization's initial contrast or starting leakage. Under separable evolution, the normalized intensity
$X_i=G_i/B_i$ is nonnegative and has expectation at most one. Independence of the sixteen realizations therefore gives
\[
 \mathbb E_{\mathrm{sep}}\left[
   \prod_{i=1}^{16}\frac{1+X_i}{2}
 \right]\leq1.
\]
Markov's inequality, followed by a Bonferroni correction across the six formation and quantum-persistence tests, gives the conservative adjusted $p$-value
\[
 p_{\mathrm{adj}}
 =\min\left\{
   1,\,
   6\left[\prod_{i=1}^{16}\frac{1+X_i}{2}\right]^{-1}
 \right\}.
\]
All realizations are included. A realization with zero target intensity contributes a factor of $1/2$.
The largest adjusted $p$-value is $1.9\times10^{-3}$. These tests use the main formation and quantum-persistence ensembles and assume the stated simulation model and independent realizations. The individual normalizations used here differ from the ratio of mean intensity to mean bound displayed in Fig.~\ref{fig:geometry-maintenance}(d).

\begingroup
\small
\bibliographystyle{unsrtnat}
\bibliography{references}

@article{Turing1952,
  author  = {Turing, A. M.},
  title   = {The Chemical Basis of Morphogenesis},
  journal = {Philosophical Transactions of the Royal Society of London. Series B, Biological Sciences},
  volume  = {237},
  number  = {641},
  pages   = {37--72},
  year    = {1952},
  doi     = {10.1098/rstb.1952.0012}
}

@article{osti_4158113,
  author = "Zeldovich, Ya. B.",
    title = "{Gravitational instability: An Approximate theory for large density perturbations}",
    journal = "Astron. Astrophys.",
    volume = "5",
    pages = "84--89",
    year = "1970"
}

@article{Polarski:1995jg,
    author = "Polarski, David and Starobinsky, Alexei A.",
    title = "{Semiclassicality and decoherence of cosmological perturbations}",
    eprint = "gr-qc/9504030",
    archivePrefix = "arXiv",
    reportNumber = "LMPM-95-4",
    doi = "10.1088/0264-9381/13/3/006",
    journal = "Class. Quant. Grav.",
    volume = "13",
    pages = "377--392",
    year = "1996"
}

@article{MUKHANOV1992203,
title = {Theory of cosmological perturbations},
journal = {Physics Reports},
volume = {215},
number = {5-6},
pages = {203-333},
year = {1992},
issn = {0370-1573},
doi = {https://doi.org/10.1016/0370-1573(92)90044-Z},
url = {https://www.sciencedirect.com/science/article/pii/037015739290044Z},
author = {V.F. Mukhanov and H.A. Feldman and R.H. Brandenberger}
}

@Article{Aschwanden2018,
author={Aschwanden, Markus J.
and Scholkmann, Felix
and B{\'e}thune, William
and Schmutz, Werner
and Abramenko, Valentina
and Cheung, Mark C. M.
and M{\"u}ller, Daniel
and Benz, Arnold
and Chernov, Guennadi
and Kritsuk, Alexei G.
and Scargle, Jeffrey D.
and Melatos, Andrew
and Wagoner, Robert V.
and Trimble, Virginia
and Green, William H.},
title={Order out of Randomness: Self-Organization Processes in Astrophysics},
journal={Space Science Reviews},
year={2018},
month={Mar},
day={06},
volume={214},
number={2},
pages={55},
issn={1572-9672},
doi={10.1007/s11214-018-0489-2},
url={https://doi.org/10.1007/s11214-018-0489-2}
}

@Article{Bond1996,
author={Bond, J. Richard
and Kofman, Lev
and Pogosyan, Dmitry},
title={How filaments of galaxies are woven into the cosmic web},
journal={Nature},
year={1996},
month={Apr},
day={01},
volume={380},
number={6575},
pages={603-606},
issn={1476-4687},
doi={10.1038/380603a0},
url={https://doi.org/10.1038/380603a0}
}

@Article{Reiter2017,
author={Reiter, F.
and S{\o}rensen, A. S.
and Zoller, P.
and Muschik, C. A.},
title={Dissipative quantum error correction and application to quantum sensing with trapped ions},
journal={Nature Communications},
year={2017},
month={Nov},
day={28},
volume={8},
number={1},
pages={1822},
issn={2041-1723},
doi={10.1038/s41467-017-01895-5},
url={https://doi.org/10.1038/s41467-017-01895-5}
}

@Article{Shankar2013,
author={Shankar, S.
and Hatridge, M.
and Leghtas, Z.
and Sliwa, K. M.
and Narla, A.
and Vool, U.
and Girvin, S. M.
and Frunzio, L.
and Mirrahimi, M.
and Devoret, M. H.},
title={Autonomously stabilized entanglement between two superconducting quantum bits},
journal={Nature},
year={2013},
month={Dec},
day={01},
volume={504},
number={7480},
pages={419-422},
issn={1476-4687},
doi={10.1038/nature12802},
url={https://doi.org/10.1038/nature12802}
}

@Article{Verstraete2009,
author={Verstraete, Frank
and Wolf, Michael M.
and Ignacio Cirac, J.},
title={Quantum computation and quantum-state engineering driven by dissipation},
journal={Nature Physics},
year={2009},
month={Sep},
day={01},
volume={5},
number={9},
pages={633-636},
issn={1745-2481},
doi={10.1038/nphys1342},
url={https://doi.org/10.1038/nphys1342}
}

@article{PhysRevA.78.042307,
  title = {Preparation of entangled states by quantum {Markov} processes},
  author = {Kraus, B. and B{\"u}chler, H. P. and Diehl, S. and Kantian, A. and Micheli, A. and Zoller, P.},
  journal = {Phys. Rev. A},
  volume = {78},
  issue = {4},
  pages = {042307},
  numpages = {9},
  year = {2008},
  month = {Oct},
  publisher = {American Physical Society},
  doi = {10.1103/PhysRevA.78.042307},
  url = {https://link.aps.org/doi/10.1103/PhysRevA.78.042307}
}

@article{10.1145/361179.361202,
author = {Dijkstra, Edsger W.},
title = {Self-stabilizing systems in spite of distributed control},
year = {1974},
issue_date = {Nov. 1974},
publisher = {Association for Computing Machinery},
address = {New York, NY, USA},
volume = {17},
number = {11},
issn = {0001-0782},
url = {https://doi.org/10.1145/361179.361202},
doi = {10.1145/361179.361202},
journal = {Commun. ACM},
month = nov,
pages = {643–644},
numpages = {2}
}

@book{10.7551/mitpress/6156.001.0001,
    author = {Dolev, Shlomi},
    title = {Self-Stabilization},
    publisher = {The MIT Press},
    year = {2000},
    month = {02},
    isbn = {9780262271837},
    doi = {10.7551/mitpress/6156.001.0001},
    url = {https://doi.org/10.7551/mitpress/6156.001.0001},
}

@article{CrossHohenberg1993,
  author  = {Cross, M. C. and Hohenberg, P. C.},
  title   = {Pattern Formation outside of Equilibrium},
  journal = {Reviews of Modern Physics},
  volume  = {65},
  number  = {3},
  pages   = {851--1112},
  year    = {1993},
  doi     = {10.1103/RevModPhys.65.851}
}

@article{GiererMeinhardt1972,
  author  = {Gierer, A. and Meinhardt, H.},
  title   = {A Theory of Biological Pattern Formation},
  journal = {Kybernetik},
  volume  = {12},
  number  = {1},
  pages   = {30--39},
  year    = {1972},
  doi     = {10.1007/BF00289234}
}

@article{MainiPainterChau1997,
  author  = {Maini, Philip K. and Painter, Kevin J. and Nguyen Phong Chau, Helene},
  title   = {Spatial Pattern Formation in Chemical and Biological Systems},
  journal = {Journal of the Chemical Society, Faraday Transactions},
  volume  = {93},
  number  = {20},
  pages   = {3601--3610},
  year    = {1997},
  doi     = {10.1039/A702602A}
}

@article{KondoMiura2010,
  author  = {Kondo, Shigeru and Miura, Takashi},
  title   = {Reaction--Diffusion Model as a Framework for Understanding Biological Pattern Formation},
  journal = {Science},
  volume  = {329},
  number  = {5999},
  pages   = {1616--1620},
  year    = {2010},
  doi     = {10.1126/science.1179047}
}

@article{YamaguchiYoshimotoKondo2007,
  author  = {Yamaguchi, Motoomi and Yoshimoto, Eiichi and Kondo, Shigeru},
  title   = {Pattern Regulation in the Stripe of Zebrafish Suggests an Underlying Dynamic and Autonomous Mechanism},
  journal = {Proceedings of the National Academy of Sciences},
  volume  = {104},
  number  = {12},
  pages   = {4790--4793},
  year    = {2007},
  doi     = {10.1073/pnas.0607790104}
}

@article{BandyopadhyayKhatunBanerjee2021,
  author  = {Bandyopadhyay, Biswabibek and Khatun, Taniya and Banerjee, Tanmoy},
  title   = {Quantum {Turing} Bifurcation: Transition from Quantum Amplitude Death to Quantum Oscillation Death},
  journal = {Physical Review E},
  volume  = {104},
  number  = {2},
  pages   = {024214},
  year    = {2021},
  doi     = {10.1103/PhysRevE.104.024214}
}

@article{KatoNakao2022,
  author  = {Kato, Yuzuru and Nakao, Hiroya},
  title   = {{Turing} Instability in Quantum Activator--Inhibitor Systems},
  journal = {Scientific Reports},
  volume  = {12},
  pages   = {15573},
  year    = {2022},
  doi     = {10.1038/s41598-022-19010-0}
}

@misc{ComparatoGarganoLoFranco2026,
  author        = {Comparato, Giorgia and Gargano, Francesco and Lo Franco, Rosario},
  title         = {{Turing} Mechanisms in a Multimode Open Quantum System},
  year          = {2026},
  eprint        = {2607.07449},
  archivePrefix = {arXiv},
  primaryClass  = {quant-ph},
  doi           = {10.48550/arXiv.2607.07449},
  note          = {arXiv:2607.07449}
}

@article{OuyangSwinney1991,
  author = {Ouyang, Qi and Swinney, Harry L.},
  title = {Transition from a uniform state to hexagonal and striped {Turing} patterns},
  journal = {Nature}, volume = {352}, pages = {610--612}, year = {1991},
  doi = {10.1038/352610a0}
}

@article{KrauseNearEquilibrium2021,
  author = {Krause, Andrew L. and Gaffney, Eamonn A. and Maini, Philip K. and Klika, V{\'a}clav},
  title = {Modern perspectives on near-equilibrium analysis of {Turing} systems},
  journal = {Philosophical Transactions of the Royal Society A: Mathematical, Physical and Engineering Sciences},
  volume = {379}, number = {2213}, pages = {20200268}, year = {2021},
  doi = {10.1098/rsta.2020.0268}
}

@article{PrigogineLefever1968,
  author = {Prigogine, Ilya and Lefever, Ren{\'e}},
  title = {Symmetry Breaking Instabilities in Dissipative Systems. {II}},
  journal = {The Journal of Chemical Physics}, volume = {48}, number = {4},
  pages = {1695--1700}, year = {1968}, doi = {10.1063/1.1668896}
}

@article{Haken1975,
  author = {Haken, Hermann},
  title = {Cooperative phenomena in systems far from thermal equilibrium and in nonphysical systems},
  journal = {Reviews of Modern Physics}, volume = {47}, number = {1},
  pages = {67--121}, year = {1975}, doi = {10.1103/RevModPhys.47.67}
}

@article{LugiatoLefever1987,
  author = {Lugiato, Luigi A. and Lefever, Ren{\'e}},
  title = {Spatial Dissipative Structures in Passive Optical Systems},
  journal = {Physical Review Letters}, volume = {58}, pages = {2209--2211},
  year = {1987}, doi = {10.1103/PhysRevLett.58.2209}
}

@article{DomokosRitsch2002,
  author = {Domokos, Peter and Ritsch, Helmut},
  title = {Collective Cooling and Self-Organization of Atoms in a Cavity},
  journal = {Physical Review Letters}, volume = {89}, pages = {253003},
  year = {2002}, doi = {10.1103/PhysRevLett.89.253003}
}

@article{BaumannEtAl2010,
  author = {Baumann, Kristian and Guerlin, Christine and Brennecke, Ferdinand and Esslinger, Tilman},
  title = {{Dicke} quantum phase transition with a superfluid gas in an optical cavity},
  journal = {Nature}, volume = {464}, pages = {1301--1306}, year = {2010},
  doi = {10.1038/nature09009}
}

@article{paul,
  author = {Paul, B. and Bandyopadhyay, Biswabibek and Banerjee, Tanmoy},
  title = {Attractive-repulsive interaction in coupled quantum oscillators},
  journal = {Physical Review E}, volume = {110}, pages = {034210}, year = {2024},
  doi = {10.1103/PhysRevE.110.034210}
}

@article{maschler,
  author = {Maschler, C. and Ritsch, Helmut and Vukics, A. and Domokos, Peter},
  title = {Entanglement assisted fast reordering of atoms in an optical lattice within a cavity at {$T=0$}},
  journal = {Optics Communications}, volume = {273}, pages = {446--450}, year = {2007},
  doi = {10.1016/j.optcom.2007.01.069}
}

@article{gatti,
  author = {Gatti, Alessandra and Mancini, Stefano},
  title = {Spatial correlations in hexagons generated via a {Kerr} nonlinearity},
  journal = {Physical Review A}, volume = {65}, pages = {013816}, year = {2001},
  doi = {10.1103/PhysRevA.65.013816}
}

@article{zhang,
  author = {Zhang, Z. and Yao, K.-X. and Feng, Lei and Hu, Jiazhong and Chin, Cheng},
  title = {Pattern formation in a driven {Bose--Einstein} condensate},
  journal = {Nature Physics}, volume = {16}, pages = {652--656}, year = {2020},
  doi = {10.1038/s41567-020-0839-3}
}

@article{liebster,
  author = {Liebster, Nikolas and Sparn, Marius and Kath, Elinor and Duchene, Jelte and Fujii, Keisuke and G{\"o}rlitz, Sarah L. and Enss, Tilman and Strobel, Helmut and Oberthaler, Markus K.},
  title = {Observation of Pattern Stabilization in a Driven Superfluid},
  journal = {Physical Review X}, volume = {15}, pages = {011026}, year = {2025},
  doi = {10.1103/PhysRevX.15.011026}
}

@ARTICLE{2026arXiv260726331I,
       author = {{Ikeda}, Kazuki},
        title = "{Quantum Turing Patterns}",
      journal = {arXiv e-prints},
         year = 2026,
        month = jul,
          eid = {arXiv:2607.26331},
        pages = {arXiv:2607.26331},
          doi = {10.48550/arXiv.2607.26331},
archivePrefix = {arXiv},
       eprint = {2607.26331},
 primaryClass = {math-ph},
       adsurl = {https://ui.adsabs.harvard.edu/abs/2026arXiv260726331I}
}

@ARTICLE{2026arXiv260820151I,
       author = {{Ikeda}, Kazuki},
        title = "{A Zoology of Quantum Turing Patterns}",
      journal = {arXiv e-prints},
         year = 2026,
        month = aug,
          eid = {arXiv:2608.20151},
        pages = {arXiv:2608.20151},
          doi = {10.48550/arXiv.2608.20151},
archivePrefix = {arXiv},
       eprint = {2608.20151},
 primaryClass = {quant-ph},
       adsurl = {https://ui.adsabs.harvard.edu/abs/2026arXiv260820151I}
}

@article{swift,
  author = {Swift, J. and Hohenberg, P. C.},
  title = {Hydrodynamic fluctuations at the convective instability},
  journal = {Physical Review A}, volume = {15}, pages = {319--328}, year = {1977},
  doi = {10.1103/PhysRevA.15.319}
}

@article{peres,
  author = {Peres, Asher}, title = {Separability criterion for density matrices},
  journal = {Physical Review Letters}, volume = {77}, pages = {1413--1415}, year = {1996},
  doi = {10.1103/PhysRevLett.77.1413}
}

@article{gillespie,
  author = {Gillespie, Daniel T.}, title = {Exact stochastic simulation of coupled chemical reactions},
  journal = {The Journal of Physical Chemistry}, volume = {81}, pages = {2340--2361}, year = {1977},
  doi = {10.1021/j100540a008}
}
\endgroup

\end{document}